\documentstyle[aps,manuscript]{revtex}

\begin{document}
\title{Quantum state of two trapped Bose-Einstein condensates with a Josephson
coupling}
\author{M. J. Steel and M. J. Collett}
\address{Department of Physics, University of Auckland, Private Bag 92 019, Auckland,%
\\
New Zealand.}
\date{14 October 1997}
\maketitle

\begin{abstract}
We consider the precise quantum state of two trapped, coupled Bose Einstein
condensates in the two-mode approximation. We seek a representation of the
state in terms of a Wigner-like distribution on the two-mode Bloch sphere.
The problem is solved using a self-consistent rotation of the unknown state
to the south pole of the sphere. The two-mode Hamiltonian is projected onto
the harmonic oscillator phase plane, where it can be solved by standard
techniques. Our results show how the number of atoms in each trap and the
squeezing in the number difference depend on the physical parameters.
Considering negative scattering lengths, we show that there is a regime of
squeezing in the relative phase of the condensates which occurs for weaker
interactions than the superposition states found by Cirac {\em et al }%
(quant-ph/9706034, 13 June 1997). The phase squeezing is also apparent in
mildly asymmetric trap configurations.
\end{abstract}

\newpage

\section{Introduction}

Traditionally, a Bose-Einstein condensate (BEC) is often viewed as a
coherent state of the atomic field with a definite phase~\cite{bar96}. It is
well known that there are problems with this view, however, related to the
fact that the phase of the atomic field is not an observable~\cite
{bar96,leg91,leg95,wis97}. The Hamiltonian for the atomic field is
independent of the condensate phase and so the correct coherent state is
only defined up to its mean number. Often it is convenient to invoke a
symmetry breaking Bogoliubov field to select a particular phase, but this
does not correspond to any physical field so the procedure is not totally
satisfactory in a formal sense. In addition, a coherent state implies a
superposition of number states, whereas in the current single trap
experiments~\cite{and95,dav95,ens96,bra97} there is a fixed number of atoms
in the trap (even if we are ignorant of that number), and the state of a
single trapped condensate must be a number state (or more precisely, a
mixture of number states). Both these problems are bypassed by considering a
system of two condensates for which the total number of atoms $N$ is fixed.
Then, a general state of the system is a superposition of number difference
states of the form 
\begin{equation}
\left| \ \right\rangle =\sum_{k=0}^{N}c_{k}\left| \ k,N-k\right\rangle .
\label{eq:expansion}
\end{equation}
As we now have a well-defined superposition state, we can legitimately
consider the relative phase of the two condensates, which is an Hermitian
observable. Indeed, the dramatic observation of interference between two
coherent BEC's~\cite{and97} constitutes a measurement of exactly this. In
the absence of atomic collisions, the expansion coefficients in Eq.~(\ref
{eq:expansion}) obey a binomial rather than Poissonian distribution as would
be expected for a coherent state.

However, there is a more straightforward objection to the identification of
the condensate with a coherent state. This is that in real experiments, the
atoms experience collisions introducing a nonlinearity into the Hamiltonian
for the system. We then know immediately that (unless very strongly damped)
the true state can not be a coherent state, leaving aside issues of absolute
versus relative phase. In the first treatment of this question, Lewenstein
and You~\cite{lew96} suggested the condensate is actually in an amplitude
quadrature eigenstate. In a fuller analysis, Dunningham ${\em et}$ {\em al~}%
\cite{dun97,dun97a} have shown that for positive (repulsive) interactions
the state is strongly number squeezed, and resembles a bent version of the
amplitude squeezed state that minimizes number fluctuations. This has the
potentially observable consequence of increasing the revival time in
collapses and revivals of the relative phase~\cite{won97,wri96,won96,wri97}
due to the reduced number variance of the squeezed state. The approach of
Dunningham {\em et al }is based on the symmetry breaking picture described
above. Their model thus describes the quantum state of a single damped
driven condensate with the phase determined by some much larger reference
condensate which does not appear in the calculation. Thus while the number
squeezing they predict is intuitively natural, the model faces the same
formal difficulties mentioned above in relation to symmetry breaking.

In this paper, we combine these two ideas by seeking an accurate description
of the ground state beyond the coherent state picture, for a system of two
coupled condensates with a fixed total number of atoms. We do this by
reducing the full quantum field theoretical description to an approximate
two-mode problem, valid for condensates of a few thousand atoms. The problem
is then well-defined in the senses discussed above: we deal with relative
rather than absolute phases, and are able to consider a completely closed
system without the complications of driving and damping, so that the ground
state is unambiguously a pure state. Using a variational approach, we then
find approximate solutions to the two-mode problem which are natural
analogues of the single condensate states found by Dunningham {\em et al.}
Our approach also works for negative (attractive) interactions and we
predict a regime of significant phase squeezing in between the coherent
state-like behavior with no interactions, and the Schr\"{o}dinger cat states
reported previously \cite{cir97,ruo97a} that occur with significant
interactions.

The paper is structured as follows. In section~\ref{sec:setup} we briefly
summarize the quantum field theory for the two-condensate problem and derive
the approximate two-mode Hamiltonian. In section~\ref{sec:outline} we
discuss a representation of the two mode states using the Bloch sphere and
outline our method for finding the ground state of the system. We construct
the solution in detail in section~\ref{sec:nonlinear}. In section~\ref
{sec:results}, we present our results and compare the predictions of our
method with exact solutions for systems with small numbers of atoms. We
consider negative scattering lengths and the associated phase squeezing in
section~\ref{sec:negchi} before we conclude.

\section{The governing Hamiltonian}

\label{sec:setup}

\subsection{Reduction to two-mode Hamiltonian}

Our model describes two condensates of atoms of mass $m$, with a linear
Josephson coupling and weak nonlinear interactions. We consider a single
trap with the condensates distinguished by their internal atomic state{\em ~}%
\cite{cir97}. The coupling is provided by laser-induced Raman transitions
between the two atomic states. Following Cirac {\em et al~}\cite{cir97}, the
second quantized Hamiltonian takes the form 
\begin{equation}
H=H_{1}+H_{2}+H_{\text{int}}+H_{\text{coup}},  \label{eq:hamiltonian}
\end{equation}
with 
\begin{eqnarray}
H_{j} &=&\int \text{d}^{3}{\bf x}\,\hat{\psi}_{j}^{\dag }\left( {\bf x}%
\right) \left[ -\frac{\hbar ^{2}}{2m}\nabla ^{2}+V_{j}\left( {\bf x}\right) +%
\frac{4\pi \hbar ^{2}a_{j}}{2m}\hat{\psi}_{j}^{\dag }\left( {\bf x}\right) 
\hat{\psi}_{j}\left( {\bf x}\right) \right] \hat{\psi}_{j}\left( {\bf x}%
\right) , \\
H_{\text{int}} &=&\frac{4\pi \hbar ^{2}a_{12}}{2m}\int \text{d}^{3}{\bf x}\,%
\hat{\psi}_{1}^{\dag }\left( {\bf x}\right) \hat{\psi}_{2}^{\dag }\left( 
{\bf x}\right) \hat{\psi}_{1}\left( {\bf x}\right) \hat{\psi}_{2}\left( {\bf %
x}\right) , \\
H_{\text{coup}} &=&-\frac{\hbar \Omega }{2}\int \text{d}^{3}{\bf x}\,\left[ 
\hat{\psi}_{1}\left( {\bf x}\right) \hat{\psi}_{2}^{\dag }\left( {\bf x}%
\right) \text{e}^{-i\delta t}+\hat{\psi}_{1}^{\dag }\left( {\bf x}\right) 
\hat{\psi}_{2}\left( {\bf x}\right) \text{e}^{i\delta t}\right] ,
\end{eqnarray}
where $j=1,$ 2. Here, the field operators $\hat{\psi}_{1}\left( {\bf x}%
\right) $ and $\hat{\psi}_{2}\left( {\bf x}\right) $ annihilate atoms at
position ${\bf x}$ in condensates $1$ and $2$ respectively, and satisfy the
relation $\left[ \hat{\psi}_{i}\left( {\bf x}\right) ,\hat{\psi}%
_{j}^{\dagger }\left( {\bf x}^{\prime }\right) \right] =\delta _{ij}\delta
\left( {\bf x}-{\bf x}^{\prime }\right) $. The term $H_{1,2}$ describes each
of the condensates in the absence of interactions with the other. They
experience spherical harmonic trap potentials $V_{1,2}$ of frequency $\omega
_{1}$ and $\omega _{2},$ and have scattering lengths $a_{1}$ and $a_{2}$
respectively. The cross-phase modulation term $H_{\text{int}}$ describes
collisional interactions between the condensates with scattering length $%
a_{12}$. The laser induced coupling is described by $H_{\text{coup}}$ with $%
\Omega $ the Rabi frequency and $\delta $ the detuning of the classical
laser field. In our work, we assume equal scattering lengths $%
a_{1}=a_{2}=a_{12},$ but allow the trap frequencies to differ.

The procedure to obtain the two-mode Hamiltonian is well-known~\cite
{cir97,mil97}. We approximate the field operators as $\hat{\psi}_{1}\left( 
{\bf x}\right) =b_{1}\phi _{1}\left( {\bf x}\right) $ and $\hat{\psi}%
_{2}\left( {\bf x}\right) =b_{2}\phi _{2}\left( {\bf x}\right) $ where $\phi
_{1,2}\left( {\bf x}\right) $ are (real) normalized mode functions for the
two condensates, and $b_{1,2}$ are the associated mode annihilation
operators which obey the standard commutation relations $\left[
b_{i},b_{j}\right] =0,$ $\left[ b_{i},b_{j}^{\dag }\right] =\delta _{ij}$.
Then Eq.~(\ref{eq:hamiltonian}) becomes

\begin{equation}
H\approx \left( \bar{\omega}_{1}+\delta \right) b_{1}^{\dagger }b_{1}+\bar{%
\omega}_{2}b_{2}^{\dagger }b_{2}+\chi _{1}b_{1}^{\dagger }b_{1}^{\dagger
}b_{1}b_{1}+\chi _{2}b_{2}^{\dagger }b_{2}^{\dagger }b_{2}b_{2}+\chi
_{12}b_{1}^{\dagger }b_{1}b_{2}^{\dagger }b_{2}-\frac{\eta }{2}\left(
b_{1}b_{2}^{\dagger }+b_{1}^{\dagger }b_{2}\right) ,  \label{eq:twomode}
\end{equation}
where 
\begin{eqnarray}
\bar{\omega}_{i} &=&\int \text{d}^{3}{\bf r}\,\bar{\phi}_{i}\left( {\bf r}%
\right) \left[ -\frac{1}{2}\nabla ^{2}+\frac{\lambda _{i}^{2}r^{2}}{2}%
\right] \bar{\phi}_{i}\left( {\bf r}\right) , \\
\chi _{i} &=&\frac{U_{i}}{2}\int \text{d}^{3}{\bf r}\,\left| \bar{\phi}%
_{i}\left( {\bf r}\right) \right| ^{4}, \\
\chi _{12} &=&\frac{U_{12}}{2}\int \text{d}^{3}{\bf r}\,\left| \bar{\phi}%
_{1}\left( {\bf r}\right) \right| ^{2}\left| \bar{\phi}_{2}\left( {\bf r}%
\right) \right| ^{2}, \\
\eta &=&\frac{\Omega }{\omega _{0}}\int \text{d}^{3}{\bf r}\,\bar{\phi}%
_{1}\left( {\bf r}\right) \bar{\phi}_{2}\left( {\bf r}\right) ,
\end{eqnarray}
for $i=1,2.$ Here we have moved into the interaction picture and introduced
dimensionless variables, scaling the Hamiltonian by an appropriate energy $%
\hbar \omega _{0}$ and the position by the length scale $x_{0}=\sqrt{\hbar
/m\omega _{0}}$ such that ${\bf r=x}/x_{0}$ and $\bar{\phi}_{i}\left( {\bf r}%
\right) =x_{0}^{3/2}\phi \left( {\bf x}\right) $. Further, $\lambda
_{i}=\omega _{i}/\omega _{0}$ and $U_{i}=4\pi a_{i}/x_{0},$ $U_{12}=4\pi
a_{12}/x_{0}.$

As shown by Milburn {\em et al}~\cite{mil97}, the same two mode model also
describes coupling between condensates in a double well potential. In this
case, the two lowest modes are strictly speaking the symmetric and
antisymmetric modes of the entire double well, but for a large dividing
potential barrier it is an accurate approximation to use modes describing
atoms in one or the other trap. The linear coupling is provided directly by
spatial tunnelling through the barrier and has a strength $\eta =\Delta
E/\omega _{0}$ where $\Delta E$ is the frequency separation of the two
(linear) modes\cite{mil97}. Assuming the potential barrier is relatively
strong, the modes are well separated and we can neglect $H_{\text{int}}$.

Equation\thinspace (\ref{eq:twomode}) defines our problem completely. For
large condensates, the mode functions are altered by the collisional
interactions and the two-mode approximation breaks down. As shown in Ref.~%
\cite{mil97}, a simple estimate shows this occurs when the number of atoms $%
N $, satisfies $Na\gg x_{0},$ where $a$ is a typical scattering length and $%
x_{0}$ is a measure of the trap size. Assuming $a\approx 5~$nm~\cite
{dav95,jin96} and a large trap with $x_{0}\approx 10$~$\mu $m, we find the
two-mode approximation should be acceptable for $N<{}2000.$

\subsection{Angular momentum representation}

The Hamiltonian can be reduced to a simpler form by exploiting the
equivalence between the algebra for two harmonic oscillators and that for
angular momentum, by introducing the new operators~\cite{sak94,mil97} 
\begin{eqnarray}
J_{+} &=&b_{1}^{\dagger }b_{2},  \nonumber \\
J_{-} &=&b_{1}b_{2}^{\dagger }, \\
J_{z} &=&\frac{1}{2}\left( b_{1}^{\dagger }b_{1}-b_{2}^{\dagger
}b_{2}\right) ,  \nonumber
\end{eqnarray}
and 
\begin{eqnarray}
J_{x} &=&\frac{1}{2}\left( J_{+}+J_{-}\right) , \\
J_{y} &=&\frac{1}{2i}\left( J_{+}-J_{-}\right) ,
\end{eqnarray}
These operators do indeed satisfy the usual angular momentum commutation
relations, justifying the choice of notation. In addition we find 
\begin{equation}
{\bf J}^{2}=J_{x}^{2}+J_{y}^{2}+J_{z}^{2}=\frac{N}{2}\left( \frac{N}{2}%
+1\right)
\end{equation}
where the total number of atoms $N=n_{1}+n_{2}=b_{1}^{\dagger
}b_{1}+b_{2}^{\dagger }b_{2}$ is a constant of the motion, so we deduce that
we are working with the angular momentum algebra for $J=N/2.$ In terms of
the new variables, the Eq.~(\ref{eq:twomode}) takes the form 
\begin{eqnarray}
H &=&J\left( \bar{\omega}_{1}+\delta +\bar{\omega}_{2}-\chi _{1}-\chi
_{2}\right) +J^{2}\left( \chi _{1}+\chi _{2}+\chi _{12}\right) +\Delta \bar{%
\omega}J_{z}+\chi _{+}J_{z}^{2}-\frac{\eta }{2}\left( J_{+}+J_{-}\right) 
\nonumber \\
&=&\Delta \bar{\omega}J_{z}+\chi _{+}J_{z}^{2}-\eta J_{x},  \label{eq:Jhamil}
\end{eqnarray}
where in the last line we have dropped an unimportant constant, and we have
introduced the effective detuning 
\begin{equation}
\Delta \bar{\omega}=\bar{\omega}_{1}+\delta -\bar{\omega}_{2}+\left(
2J-1\right) \chi _{-},
\end{equation}
and effective nonlinearity 
\begin{equation}
\chi _{+}=\chi _{1}+\chi _{2}-\chi _{12}.
\end{equation}
Note the useful fact that both the difference in the self-nonlinearities $%
\chi _{-}=\chi _{1}-\chi _{2}$ and the cross-nonlinearity $\chi _{12}$
merely shift the values of $\Delta \bar{\omega}$ and $\chi _{+}$ and
introduce no new terms. Thus there is no restriction of the physics by
assuming equal scattering lengths. We now calculate these parameters for
realistic experimental values. Taking the Na$^{23}$ atom for example, we
have $a\approx 5$~nm, and suppose trap frequencies of order $\omega
_{i}=1000 $~s$^{-1}$\cite{dav95}$.$ Then taking the scaling frequency $%
\omega _{0}=1$~s$^{-1},$ we obtain $\bar{\omega}_{i}\approx 1500,$ and $\chi
_{i}\approx 1.4. $ Therefore, the detuning $\Delta \bar{\omega}$ may range
from zero to a few hundred. The coupling strength $\eta $ is largely
arbitrary. In the spatial case, it can take values up to the order of the
trap frequencies~\cite{mil97}, or can be made as small as desired by
incresing the trap separation.

\section{Outline of approach}

\label{sec:outline}

For the remainder of the paper we are concerned with the ground state of
Eq.~(\ref{eq:Jhamil}). Milburn {\em et al}~\cite{mil97} have presented
numerical calculations of the energy spectrum of this Hamiltonian and the
dynamical problem has also been studied~\cite{mil97,kor97}. Rather than the
spectrum or dynamics, however, our concern is with the detailed properties
of the lowest eigenstate and their dependence on the effective detuning and
nonlinearity. In general, the eigenstates of Eq.~(\ref{eq:Jhamil}) can not
be written analytically. For systems with at most a few hundred atoms, it is
feasible to find the exact eigenstates in the basis of $J_{z}$ eigenstates $%
\left| J,m=-J,\ldots ,J\right\rangle _{z}$ numerically. Our semi-analytic
approach can be used for systems of arbitrary size and lends considerable
insight to the problem.

\subsection{Bloch sphere}

\label{sec:blochsphere}

Our approach relies closely on the Bloch sphere representation of angular
momentum which we must briefly introduce. A detailed analysis has been given
by Arecchi {\em et al~}\cite{are72}. Quantum states in the angular momentum
Hilbert space can be usefully represented on the Bloch sphere. Certain
states---the atomic coherent states or ``Bloch'' states~\cite{are72}%
---correspond to a single point on the sphere. Defined as the rotated states 
$|\theta ,\varphi \rangle =R_{\theta ,\varphi }|J,-J\rangle _{z},$ where the
rotation operator 
\begin{equation}
R_{\theta ,\varphi }=\exp \left[ \theta /2\left( J_{+}\text{e}^{-i\phi
}-J_{-}\text{e}^{i\phi }\right) \right] =\exp \left[ -i\theta \left(
J_{x}\sin \varphi -J_{y}\cos \varphi \right) \right] ,  \label{eq:rotop}
\end{equation}
they are labeled by the spherical coordinates $\theta $ and $\varphi $
corresponding to the state's point on the sphere. Note that in terms of our
BEC problem, the north pole $\left| \theta =\pi \right\rangle $ and south
pole $\ \left| \theta =0\right\rangle $ represent the states with all atoms
in mode 1 or 2 respectively. States lying on the equator with $\theta =\pi
/2 $ represent an equal division of atoms between the modes, (which for $%
\eta \ne 0,$ does {\em not }imply the number state $\left|
N/2,N/2\right\rangle ,$ but rather an entanglement of the form~(\ref
{eq:expansion}) with a binomial distribution of expansion coe\-\-fficients).
The Bloch states are the analogs in the angular momentum algebra of the
standard coherent states of the harmonic oscillator{\em ~}\cite{are72}. They
share a number of properties with the coherent states, for instance minimum
uncertainty in the natural variables. In addition, more general
non-classical states described by the state vector $\ \left| \psi
\right\rangle $ or density matrix $\rho $ can be naturally pictured in terms
of a quasi-probability distribution function on the sphere 
\begin{eqnarray}
\tilde{Q}_{\ \left| \psi \right\rangle }\left( \theta ,\varphi \right)
&=&\left| \left\langle \theta ,\varphi \left| \psi \right. \right\rangle
\right| ^{2}, \\
\tilde{Q}_{\rho }\left( \theta ,\varphi \right) &=&\left\langle \ \theta
,\varphi \right| \rho \left| \theta ,\varphi \right\rangle ,
\end{eqnarray}
in analogy to the familiar $Q$ function in the harmonic oscillator
phase-plane~\cite{wal94}. Functions analogous to the standard
Glauber-Sudarshan $P$ and Wigner distributions can also be defined. As
discussed below, these analogies can be made precise using a formal
contraction from the angular momentum Hilbert space to the Hilbert space for
a single harmonic oscillator~\cite{are72}. While the Bloch states lack some
of the useful properties of the coherent states~\cite{are72,twotrapendnote1}%
, nonetheless, we see below that the Wigner or $\tilde{Q}$ functions make
for useful measures of quantities such as the squeezing in the number
difference or relative phase in the ground state.

\subsection{Mathematical procedure}

The angular momentum commutation relations make a direct solution to our
problem in the full Hilbert space difficult. The previous section suggests
the following alternative approach. We assume the ground state we seek has a
quasi-probability distribution localized to a particular part of the Bloch
sphere. (Thus we immediately exclude Schr\"{o}dinger cat states such as
those found by Cirac {\em et al~}\cite{cir97} and Ruostekoski {\em et al}~%
\cite{ruo97a}, which we treat numerically in section~\ref{sec:negchi}.) We
apply a rotation to the Hamiltonian to bring the mean value of the state to
the south pole of the Bloch sphere. This must be done self-consistently as
we do not actually know the mean value of the state until we have solved the
problem. We then project the problem to the harmonic oscillator phase plane
using the contraction operation of Ref.~\cite{are72}. The problem can then
be solved in the plane and the ground state plotted as a Wigner
distribution. Finally, we project this distribution back on to the sphere,
and rotate it to the original mean value which has by now been determined.
Equivalently, we can think of the problem being solved in the oscillator
phase plane that is tangent to the sphere at the mean value of the state. In
section~\ref{sec:nonlinear} we make these ideas precise and provide the
solution.

\subsection{Linear problem}

In the absence of the nonlinear term ($\chi _{+}=0$), the problem is
trivial. The Hamiltonian~(\ref{eq:Jhamil}) becomes 
\begin{equation}
H=\Delta \bar{\omega}J_{z}-\eta J_{x},
\end{equation}
and using the rotation relations in appendix~\ref{appen:nonlin},{\em \ }it
is easy to see that the rotated Hamiltonian $H^{\prime }=R_{\theta ,\pi
}HR_{\theta ,\pi }^{-1}$ where $\tan \theta =\eta /\Delta \bar{\omega},$ is
just a multiple of $J_{z}$ with ground state $\left| J,-J\right\rangle _{z}.$
Inverting the rotation, the exact ground state for the original Hamiltonian
is simply the Bloch state $\left| \theta ,\phi \right\rangle =\left| \tan
^{-1}\left( \eta /\Delta \bar{\omega}\right) ,0\right\rangle .$ Thus the the
ground state of coupled ideal gas condensates is the entangled state analog
of the coherent state. This result is well-known, though it is more commonly
expressed in the number difference basis~\cite{mol97}. As expected, for
vanishing $\Delta \bar{\omega}$, the traps are equivalent and there is an
equal number of atoms in each, whereas for $\Delta \bar{\omega}\ne 0,$ the
ground state has more atoms in the weaker trap. Note that the relative phase
of the condensates $\varphi =0.$ This is the reason for our choice of the
negative sign in front of $\eta $ in the original Hamiltonian (\ref
{eq:Jhamil}). Even when we consider the more complicated states of the
nonlinear system, we see by symmetry that the mean value of the state must
still have $\varphi =0,$ simplifying the rotation operators we need to
consider. In practice, the phase of $\eta $ is determined by the phase of
the driving laser field. By a suitable choice of coordinates we may always
take it to be zero.

\section{Nonlinear problem}

\label{sec:nonlinear}

\subsection{Contraction from angular momentum to harmonic oscillator Hilbert
space}

The nonlinear problem is much more involved. As indicated earlier, the first
step is to perform a rotation of the Hamiltonian by an undertermined angle $%
\theta $ and then project the new Hamiltonian into the harmonic oscillator
phase plane with operators $a,a^{\dag }$ satisfying $\left[ a,a^{\dag
}\right] =1.$ This procedure can be made rigorous through the concept of a
group contraction from the angular momentum Hilbert space to the harmonic
oscillator Hilbert space. The details can be found in Ref.~\cite{are72} to
which we refer the interested reader. Quoting the results, the contraction
is made by the identification of operators according to 
\begin{eqnarray}
J_{+} &\rightarrow &\frac{1}{c}a^{\dag },  \label{eq:mapping1} \\
J_{-} &\rightarrow &\frac{1}{c}a, \\
J_{z} &\rightarrow &a^{\dag }a-\frac{1}{2c^{2}},
\end{eqnarray}
where $c=1/\sqrt{2J}$ and the spaces are formally identical in the limit $%
c\rightarrow 0$. In the same limit$,$ we can contract eigenstates of $J_{z}$
to the harmonic oscillator number states: 
\begin{equation}
\ \left| J,M\right\rangle \rightarrow \ \left| n=J+M\right\rangle ,
\end{equation}
and we relate the coordinates according to 
\begin{equation}
\frac{\theta }{2}\exp \left( i\varphi \right) \rightarrow c\alpha .
\label{eq:mapping5}
\end{equation}
Later we also use the quadrature operators $X=a+a^{\dag }$ and $Y=-i\left(
a-a^{\dag }\right) ,$ which are the contractions of $J_{x}$ and $J_{y}$
respectively. Geometrically, we visualize this contraction as a projection
from the Bloch sphere to the phase plane with the south pole of the sphere
coincident with the origin of the phase plane. Note that the coherent
amplitude $\alpha $ can take values throughout the whole phase plane only in
the limit $c\rightarrow 0$ and there is naturally a distortion involved in
the projection.Physically, by performing the contraction we discard the
knowledge that the true ladder of states is bounded at both ends rather than
just the lower end. However, providing the state is localized near the south
pole and $c$ $\ll 1$ (large atom number), the distortion is small. The
contraction process also maps functions from the sphere to the plane, so for
instance we can identify the (rotated) Bloch sphere distribution function $%
\tilde{Q}\left( \theta ,\varphi \right) $ with the standard phase plane
function $Q\left( \alpha \right) =\left\langle \alpha \left| \rho \right|
\alpha \right\rangle $. Here, we a define the Wigner-like distribution on
the sphere by a projection of the harmonic oscillator Wigner distribution
using Eq.~(\ref{eq:mapping5}) in reverse.

\subsection{Variational solution}

\subsubsection{Gaussian part}

We are at last ready to find our approximate solution to the full problem.
The procedure is somewhat involved mathematically, and we state only the
main intermediate steps here, leaving the details to appendix~\ref
{appen:nonlin}. The first step is to rotate the Hamiltonian~(\ref{eq:Jhamil}%
) around the positive $y$-axis by an undetermined angle $\theta ,$ and
perform the contraction operation to the single oscillator phase space to
find the new Hamiltonian $F.$ Following Dunningham {\em et al~}\cite{dun97a}%
, we write this Hamiltonian as 
\begin{equation}
F=F_{G}+F_{NG},
\end{equation}
in terms of a Gaussian part $F_{G}$ and non-Gaussian part $F_{NG},$ the
latter of which satisfies the constraints 
\begin{eqnarray}
\left\langle F_{NG}\right\rangle &=&0,  \nonumber \\
\left\langle \left[ a,F_{NG}\right] \right\rangle &=&\left\langle \left[
a^{\dag },F_{NG}\right] \right\rangle =0,  \label{eq:constraints} \\
\left\langle \left[ a,\left[ a,F_{NG}\right] \right] \right\rangle
&=&\left\langle \left[ a,\left[ a^{\dag },F_{NG}\right] \right]
\right\rangle =0.  \nonumber
\end{eqnarray}
This separation allows us to find the ground state of the Gaussian part
first, and by assuming a weak nonlinearity, treat the non-Gaussian part as a
perturbation. Appendix \ref{appen:nonlin} contains the expression for the
non-Gaussian part. The Gaussian part is 
\begin{equation}
F_{G}=K+L\left( a+a^{\dag }\right) +Sa^{\dag }a+T\left( a^{2}+a^{\dag
2}\right) ,  \label{eq:Fgauss}
\end{equation}
where 
\begin{eqnarray}
K &=&\chi _{+}J/2\sin ^{2}\theta +\chi _{+}J^{2}\cos ^{2}\theta -J\left(
\Delta \bar{\omega}\cos \theta +\eta \sin \theta \right)  \nonumber \\
&&+2\sqrt{2J}\chi _{+}\cos \theta \sin \theta \ \left\langle a^{\dagger
}a^{2}\right\rangle +\chi _{+}\cos ^{2}\theta \left( \ \left\langle
a^{\dagger 2}a^{2}\right\rangle -2\ \left\langle a^{2}\right\rangle ^{2}-4\
\left\langle a^{\dagger }a\right\rangle ,^{2}\right)  \label{eq:Kdef} \\
L &=&\sqrt{\frac{J}{2}}\left\{ \Delta \bar{\omega}\sin \theta -\eta \cos
\theta +\chi _{+}\sin \theta \cos \theta \left[ -\left( 2J-1\right)
+2\left\langle a^{2}\right\rangle +4\left\langle a^{\dag }a\right\rangle
\right] \right\} ,  \label{eq:Ldef} \\
S &=&\eta \sin \theta +\Delta \bar{\omega}\cos \theta +\chi _{+}\left\{
J\sin ^{2}\theta +\cos ^{2}\theta \left[ -\left( 2J-1\right) +4\left\langle
a^{\dag }a\right\rangle \right] \right\} ,  \label{eq:Sdef} \\
T &=&\chi _{+}\left( \frac{J}{2}\sin ^{2}\theta +\cos ^{2}\theta
\left\langle a^{2}\right\rangle \right) .  \label{eq:Tdef}
\end{eqnarray}
In Eqs.~(\ref{eq:Kdef})--(\ref{eq:Tdef}), we have taken $\ \left\langle
a^{2}\right\rangle =\left\langle a^{\dag 2}\right\rangle $ which follows
from the choice of $\eta $ as real. Note that $F_{G}$ depends on moments
taken over the state which is the solution we are seeking. We can solve the
Gaussian part to different levels of accuracy according to how we account
for these expectation values.

\paragraph{Non self-consistent approach}

We first assume we have known values for the expectation values. For
example, we may take a mean-field approximation in which all the moments are
zero, or as explained below we may have obtained estimates for the moments
from a previous less accurate calculation (such as the mean-field one). We
consider a self-consistent approach in the next section. The rotation angle $%
\theta $ is fixed by requiring that the linear terms should vanish: 
\begin{equation}
L=0.  \label{eq:condition1}
\end{equation}
Except for a constant term, $F_{G}$ is now purely quadratic and we perform a
Boguliobov diagonalization by writing 
\begin{eqnarray}
a &=&b\cosh r-b^{\dag }\sinh r,  \nonumber \\
a^{\dag } &=&b^{\dag }\cosh r-b\sinh r.  \label{eq:bogul}
\end{eqnarray}
Substituting Eqs.~(\ref{eq:bogul}) into Eq.~(\ref{eq:Fgauss}) and setting
the terms in $b^{2}$ and $b^{\dag 2}$ to zero, we obtain 
\begin{equation}
\exp \left( -2r\right) =\sqrt{\frac{S-2T}{S+2T}},
\end{equation}
while the diagonalized Hamiltonian is 
\begin{equation}
F_{G}=K+(S\cosh ^{2}r-2T\cosh r\sinh r)+\sqrt{S^{2}-4T^{2}}b^{\dag }b.
\label{eq:diaghamil}
\end{equation}
The first two terms are constants, so the ground state is just the vacuum in
the $b$ representation. As the transformation~(\ref{eq:bogul}) is induced by
the squeezing operator $S\left( r\right) =\exp \left[ r\left( a^{2}-a^{\dag
2}\right) /2\right] ~$\cite{wal94}, the $b$ eigenstates $\left|
i\right\rangle $ transform back as 
\begin{equation}
\left| \tilde{i}\right\rangle _{a}=S\left( r\right) \left| i\right\rangle .
\end{equation}
Thus the ground state in the $a$ representation is just the squeezed vacuum
with 
\begin{equation}
\ \left\langle X^{2}\right\rangle =\exp \left( -2r\right) =\sqrt{\frac{S-2T}{%
S+2T}}.  \label{eq:condition2}
\end{equation}
In the mean-field limit we have the simple results 
\begin{eqnarray}
\Delta \bar{\omega} &=&\eta \cot \theta +\chi _{+}\cos \theta \left(
2J-1\right) ,  \label{eq:thetaclosed} \\
\left\langle X^{2}\right\rangle &=&\sqrt{\frac{\eta }{\eta +N\chi _{+}\sin
^{3}\theta }}.  \label{eq:X2closed}
\end{eqnarray}
For $\Delta \bar{\omega}=0,$ we find symmetric states with $\theta =\pi /2$
as is natural. Moreover, in the limit $\theta \rightarrow $ $\pi /2$ (that
is, for $\Delta \bar{\omega}/\left[ \left( 2J-1\right) \chi _{+}\right]
\rightarrow 0$), the non-Gaussian part of the Hamiltonian $F_{NG}=0$ (see
appendix~\ref{appen:nonlin}) and $F_{G}$ is independent of any expectation
values. Thus in this limit, the projected state is {\em exactly} a squeezed
state with $\left\langle X^{2}\right\rangle =\sqrt{\eta /\left( \eta +\chi
_{+}N\right) }$.We note in passing that for a negative nonlinearity$,$ Eq.~(%
\ref{eq:X2closed}) predicts $\left\langle X^{2}\right\rangle >1$ which
indicates a possibility of phase-squeezing. We return to this in section~\ref
{sec:negchi}.

\paragraph{Self-consistent Approach}

We can also find the ground state of Eq.~(\ref{eq:Fgauss}) with a
self-consistent approach in which the expectation values are determined to
Gaussian approximation in the course of the calculation. In this case, the
first two terms in Eq.~(\ref{eq:diaghamil}) can not be considered constants
and in general the $b$-vacuum is not the lowest eigenstate. The correct
approach is to assume a squeezed vacuum solution $\left| r\right\rangle
=S\left( r\right) \left| 0\right\rangle $ to Eq.~(\ref{eq:Fgauss}) and find $%
\theta $ and $r$ by minimizing the expectation value of the energy subject
to the constraint in Eq.~(\ref{eq:condition1}). Performing the
transformation~(\ref{eq:bogul}) with this value of $r,$ gives a Hamiltonian
in the $b$ representation with a small off-diagonal part which can be
transfered to the non-Gaussian part $F_{NG}$ yet to be treated. In fact, we
have found that we can obtain virtually identical results by proceeding
directly with the Boguliobov diagonalization, and solving Eqs.~(\ref
{eq:condition1}) and~(\ref{eq:condition2}) simultaneously for $\theta $ and $%
r,$ where $L,$ $S$ and $T$ now depend on $r$ through the quadratic moments.

\subsubsection{Non-Gaussian part}

We now include the effects of the non-Gaussian part $F_{NG}$ as a
perturbation to the squeezed state just found. We use second-order
perturbation theory to write the corrected state as 
\begin{eqnarray}
\left| \Phi ^{\left( 1\right) }\right\rangle _{a} &=&\left| \tilde{0}%
\right\rangle _{a}+\sum_{k\ne 0}\frac{\mbox{\raisebox{-.7ex}{$_{a}$}}%
\!\left\langle \tilde{k}\right| F_{NG}\left| \tilde{0}\right\rangle _{a}}{%
E_{0}^{\left( 0\right) }-E_{k}^{\left( 0\right) }}\left| \tilde{k}%
\right\rangle _{a} \\
&&+\sum_{k\ne 0}\left\{ \left[ \sum_{l\ne 0}\frac{\mbox{%
\raisebox{-.7ex}{$_{a}$}}\!\left\langle \tilde{k}\right| F_{NG}\left| \tilde{%
l}\right\rangle _{aa}\!\left\langle \tilde{l}\right| F_{NG}\left| \tilde{0}%
\right\rangle _{a}}{\left( E_{0}^{\left( 0\right) }-E_{k}^{\left( 0\right)
}\right) \left( E_{0}^{\left( 0\right) }-E_{l}^{\left( 0\right) }\right) }%
\right] -\frac{\mbox{\raisebox{-.7ex}{$_{a}$}}\!\left\langle \tilde{k}%
\right| F_{NG}\left| \tilde{0}\right\rangle _{aa}\!\left\langle \tilde{0}%
\right| F_{NG}\left| \tilde{0}\right\rangle _{a}}{\left( E_{0}^{\left(
0\right) }-E_{k}^{\left( 0\right) }\right) ^{2}}\right\} \left| \tilde{k}%
\right\rangle _{a}.
\end{eqnarray}
where from Eq.~(\ref{eq:diaghamil}) $E_{k}^{\left( 0\right) }=k\sqrt{%
S^{2}-4T^{2}}$ is the energy of the unperturbed state $\left| \tilde{k}%
\right\rangle _{a}$.We then calculate the Wigner distribution 
\begin{equation}
W\left( \alpha \right) =\frac{1}{\pi ^{2}}\int \text{d}^{2}z\,\text{e}%
^{\alpha z^{*}-\alpha ^{*}z}\chi \left( z\right) ,
\end{equation}
where the symmetric characteristic function is $\chi \left( z\right) =$ Tr$%
\left\{ \rho _{a}\exp \left( za^{\dag }-z^{*}a\right) \right\} $ and $\rho
_{a}=\left| \Phi ^{\left( 1\right) }\right\rangle _{aa}\left\langle \Phi
^{\left( 1\right) }\right| $. The details of these calculations are given in
appendix~\ref{appen:wigner}. The final answer has a closed form in terms of
Hermite Gaussians but is too long to write here.

\subsubsection{Combined Approach}

In finding ground states we have used the above steps in an iterative
scheme. A first approximation is found using the self-consistent approach to
the Gaussian part followed by the perturbation theory. The quadratic moments
appearing in Eqs.\ (\ref{eq:condition1}) and\ (\ref{eq:condition2}) [through 
$L,$ $S$ and $T]$ are calculated using this first approximation and a new
Gaussian state chosen. Finally the perturbation theory is applied again.

\subsubsection{Anti-rotation}

To complete the problem, the contours of the Wigner function just obtained
must be projected back to the sphere and the original rotation of the
Hamiltonian by angle $\theta $ reversed. This is completely elementary and
we reserve the equations for appendix~\ref{appen:antirotate}.

\section{Results}

\label{sec:results}

\subsection{Exact States}

In this section, we present a mixture of exact numerical results and those
obtained by our semi-analytic procedure. This allows us to test the
agreement in the regime where the size of the Hilbert space is small enough
to permit a complete numerical solution. To begin, in Figure~\ref
{fig:exactsphere}, we show three sample exact states for $N=100$ atoms
plotted as contours of Wigner functions on the surface of the Bloch sphere.
There are two contours for each state at heights e$^{-1}$ and e$^{-1}/4$ of
the maximum of the Wigner function. States (a) and (b) show states with a
nonlinearity $\chi _{+}=0.75$ and detuning $\Delta \omega =0$ and 30,
respectively. Both states show strong squeezing in the number difference
(the vertical axis $J_{z}$), while for the asymmetric case (b) the atoms are
predominantly found in the trap of lower energy. Note that the sense of
squeezing is along parallels of latitude and not along the great circle
through the mean field point. This is rather obvious---we expect squeezing
along the number difference axis $J_{z},$ but it has the effect that the
states are in most cases far from minimum uncertainty in the natural
variables. We discuss this shortly. For comparison, the state (c) is just
the Bloch state solution to the linear problem $\left( \chi _{+}=0\right) $,
with $\Delta \omega =-0.44,$ for which the contours are circles. Note that
the mean angle $\theta $ is the same distance from the equator $\theta =\pi
/2$ for states (b)\ and (c) despite very different values of the magnitude
of the detuning $\left| \Delta \omega \right| .$ As indicated by Eq.~(\ref
{eq:thetaclosed}), the positive nonlinearity tends to push states back
towards the equator and is balanced by a much larger value of the detuning.
This derives from an energy competition between the terms in $\left\langle
J_{z}\right\rangle $ and $\left\langle J_{z}^{2}\right\rangle $ in the
Hamiltonian~(\ref{eq:Jhamil}).

\subsection{Comparison with the model}

We illustrate the results of our semi-analytic method by rotating the
nonlinear states (a) and (b) in Fig.~\ref{fig:exactsphere} to the south
pole, and projecting them to the plane in Figs.~\ref{fig:exactplane}(a) and~%
\ref{fig:exactplane}(b) respectively. The contours are at e$^{-1}/2$ of the
maximum of the Wigner function. Here we see the effect of the fact that the
orientation of the squeezing is along the parallels of latitude. The state
originally at $\theta =\pi /2$ [Fig~\ref{fig:exactsphere}(a)] is a precise
squeezed state with no bending, but the asymmetric state in Fig.~\ref
{fig:exactsphere}(b) is distorted on projection [solid line in Fig.~\ref
{fig:exactplane}(b)]. The other lines in ~\ref{fig:exactplane}(b) indicate
our semi-analytic prediction to second order perturbation theory. The dashed
line shows the solution using the mean-field approximation for the
expectation values in Eqs.~(\ref{eq:Kdef})--(\ref{eq:Tdef}). The dot-dashed
line is for the improved result in which the expectation values are first
estimated using the self-consistent approach. The bending we find here is a
clear analog of that found for a single condensate by Dunningham {\em et al~}%
\cite{dun97a} but in our case arises purely from the geometric effect of
projection. As our theory gives the exact symmetric state, the lines are
coincident in Fig.\ref{fig:exactplane}(a).

The dependence of the mean angular position of the state $\theta =\tan
^{-1}\left( -\left\langle J_{x}\right\rangle /\left\langle
J_{z}\right\rangle \right) $ on the detuning and nonlinearity for exact
solutions with $N=200$ is shown in Fig.~\ref{fig:exacttheta}(a). This, of
course is a measure of the imbalance in the populations of each trap: $%
\left\langle n_{1}\right\rangle =J\left( 1-\cos \theta \right) ,$ $%
\left\langle n_{2}\right\rangle =J\left( 1+\cos \theta \right) .$ We plot
the mean angle $\theta $ as a function of the nonlinearity $\chi _{+}$ for
detunings of $\Delta \bar{\omega}$ $=0$, 5, 25, 50, 75 and 100 which label
the curves. As $\chi _{+}$ increases, the mean value increases from the
linear result $\eta =\tan ^{-1}\left( \eta /\Delta \bar{\omega}\right) $
towards the symmetric value $\theta =\pi /2$, with the curves for larger
detuning shifting at larger nonlinearities. From Eq.~(\ref{eq:thetaclosed}),
we see that the most rapid change occurs for $\chi _{+}\approx \Delta \bar{%
\omega}/\left( 2J-1\right) $. As explained above, the tendency toward
symmetric states is a result of an increasing energy penalty for asymmetric
states from the $\left\langle J_{z}^{2}\right\rangle $ term in the
Hamiltonian. We check the accuracy of our model in Fig.~\ref{fig:exacttheta}%
(b) showing the discrepancy in the mean angle $\theta $ according to the
mean-field (dotted line) and self-consistent predictions\ (solid line), from
the exact value calculated numerically. The curves are labeled with the
value of the detuning $\Delta \bar{\omega}.$ There is a clear improvement
with the self-consistent case, though it is less dramatic for the larger
detuning.

We consider the behaviour of the spread in number difference $\delta n=\sqrt{%
\text{Var}\left( n_{1}-n_{2}\right) }=4\sqrt{\text{Var}\left( J_{z}\right) }$
for the same parameters in Fig.~\ref{fig:numvar}. The solid lines are the
exact result, the dotted lines our approximate result using the corrected
quadratic moments, and again the curves are labeled by the detuning $\Delta 
\bar{\omega}$. For $\Delta \bar{\omega}$ $=0,$ the state is always centered
on the equator and the number squeezing grows stronger with the
nonlinearity. For this case, in the limit $c\rightarrow 0$ when the
projection gives the exact solution, we have $\delta n=\sqrt{N}\left[ \eta
/\left( \eta +\chi _{+}N\right) \right] ^{1/4}.$ The discrepancy of this
curve from the exact result is not visible in Fig.~\ref{fig:numvar}. The
behavior is somewhat different for the other cases. Initially the spread in
number increases, before turning around and becoming coincident with the
decreasing symmetric case. The initial rise in the variance agrees closely
with the Bloch state result $\delta n=$ $\sqrt{N\sin \theta }$ (not shown in
figure) until just before the maxima of the curves. Thus we see that
initially the nonlinearity shifts the mean value of the state without
affecting its shape. In this plot, we see that our approximate method is
less successful for cases with large detunings. These states are highly
asymmetric and therefore the projected states show significant bending. The
perturbation from the Gaussian squeezed state is thus larger and our
calculation less accurate.

\section{Negative nonlinearities}

\label{sec:negchi}We demonstrate briefly here that for a negative
nonlinearity there is a regime of phase squeezing rather than number
squeezing. Using a mean-field picture, Cirac{\em \ et al}~\cite{cir97} have
found a range of superposition states for negative nonlinearities
(attractive interactions). They show the two lowest energy states are even
and odd superpositions of states in which most of the atoms are in trap 1 or
most are in trap 2. In our notation they arise as follows. In the mean field
approximation of Eq.~(\ref{eq:thetaclosed}) and taking the symmetric case $%
\Delta \bar{\omega}=0,$ we have 
\begin{equation}
\sin \theta =\frac{-\eta }{\chi _{+}\left( N-1\right) }.  \label{eq:negmean}
\end{equation}
This equation clearly only has solutions for $\left| \chi _{+}\right| $
sufficiently large. When this is true, there are two degenerate mean field
ground states $\left| \theta ,0\right\rangle $ and $\left| \pi -\theta
,0\right\rangle .$ Cirac {\em et al} have shown that the superposition or
``Schr\"{o}dinger'' cat states $\left| \pm \right\rangle =\frac{1}{\sqrt{2}}%
\left( \left| \theta ,0\right\rangle \pm \left| \pi -\theta ,0\right\rangle
\right) $ give a lower value for the energy and thus are a better
approximation to the lowest energy levels. Ruostekoski and Walls~\cite
{ruo97a} have proposed a scheme for generating similar superpositions in
number for free condensates. Numerically, we have found that the exact
ground states in this regime are indeed of a superposition nature, though of
course they are superpositions of distorted Bloch states, not of true Bloch
states. What about the regime $\eta >\left| \chi _{+}\right| \left(
2J-1\right) $ for which Eq.~(\ref{eq:negmean}) has no solutions? Equations~(%
\ref{eq:thetaclosed}) and~(\ref{eq:X2closed}) give solutions with $\theta
=\pi /2$ and $\left\langle X^{2}\right\rangle >1.$ As in the Gaussian
approximation the states are minimum uncertainty we are led to expect phase
squeezing. While our method is applicable for negative nonlinearities, the
states can be highly non-Gaussian and the variational method is not always
very successful. Therefore we use numerical results to indicate that the
phase squeezing does indeed occur. We reduce the number of atoms to $N=100$
to make squeezing more obvious in the figures. Thus in the mean field
approximation, we expect superposition states for $\chi _{+}<-1/99\approx
-0.0101.$ In Figs.~\ref{fig:negsqueeze}(a)-(b) we show the Wigner function
for a succession of states with $\Delta \bar{\omega}=0,$ and nonlinearities
(a)~$\chi _{+}=-0.01,$ (b)~$\chi _{+}=$ $-0.0115.$ For a vanishing
nonlinearity we would have circles centered on the equator. The state
becomes increasingly elongated in the number difference direction and
strongly squeezed in the relative phase direction around the equator. Note
that state (b) lies in the range where the mean field picture predicts a cat
state. With further increase to $\chi _{+}=-0.012$, the phase squeezed state
bifurcates to the cat state. This is seen in Fig.~\ref{fig:negsqueeze}(c)
where we have plotted the $\tilde{Q}$ function rather than the Wigner
function to avoid interference fringes. In Fig.~\ref{fig:negsqueeze}(d) we
treat an asymmetric case with $\Delta \bar{\omega}=0.001,$ $\chi
_{+}=-0.0115 $. Here, the energy gained by adopting the superposition state
is outweighed by the energy difference between the two traps and the lowest
energy state is a single drawn out ``tear-drop''. The extended tail is
clearly a vestige of the superposition states that are favorable for
vanishing or very small asymmetries. The long tail and phase squeezing may
be thought of as a ``best attempt'' to attain a cat-like state. In Fig.~\ref
{fig:phasesqueeze} we show the phase variance $\Delta \phi =\left(
\left\langle J_{y}^{2}\right\rangle -\left\langle J_{y}\right\rangle
^{2}\right) /(J/2)$ as a function of the nonlinearity for several values of
the detuning $\Delta \bar{\omega}.$ For small asymmetries there is strong
phase squeezing. At larger detunings, the system is too far from the
superposition state regime and the residual phase squeezing is quenched out.

\section{Conclusion}

In this paper we have studied the quantum statistics of the ground state of
a two-mode model for coupled Bose-Einstein condensates. We find strong
squeezing of the number difference for positive nonlinearities and a regime
of squeezing in the relative phase for negative nonlinearities. Within the
validity of the two-mode approximation our model can treat systems of
arbitrary numbers of atoms. However, its applicability is limited by the
eventual distortion of the condensate mode functions that occurs for
condensates of more than a few thousand atoms. In order to treat larger
condensates, one must account for a larger number or possibly all of the
trap modes. This might be attempted by a variational solution of the full
second-quantized Hamiltonian. In this fashion, Cirac{\em \ et al~}\cite
{cir97} have calculated the energies of superposition state, while Spekkens
and Sipe~\cite{spe97} have considered the coherence properties of double
traps, but neither have discussed the detailed shape of the ground state.
Other authors are currently using stochastic simulations of generalized
Gross-Pitaevski equations with additional quantum noise terms to account for
the higher modes~\cite{ols97}.

\acknowledgments

We acknowledge support of the Marsden fund of the Royal Society of New
Zealand, the University of Auckland Research Committee and the New Zealand
Lotteries' Grants Board.

\appendix 

\section{Separation of the contracted Hamiltonian}

\label{appen:nonlin}

Here we provide a fuller account of some of the steps in finding the ground
state in the single oscillator Hilbert space. We first note that the
rotation operator~(\ref{eq:rotop}) transforms $J_{x}$ and $J_{z}$ as 
\begin{eqnarray}
R_{\theta ,\pi }J_{x}R_{\theta ,\pi }^{-1} &=&J_{x}\cos \theta -J_{z}\sin
\theta , \\
R_{\theta ,\pi }J_{z}R_{\theta ,\pi }^{-1} &=&J_{x}\sin \theta +J_{z}\cos
\theta .
\end{eqnarray}
Using these relations, we rotate the original Hamiltonian~(\ref{eq:Jhamil})
to obtain. 
\begin{eqnarray}
H^{\prime } &=&R_{\theta ,\pi }HR_{\theta ,\pi }^{-1}  \nonumber \\
&=&J_{x}\left( \Delta \bar{\omega}\sin \theta -\eta \cos \theta \right)
+J_{z}\left( \Delta \bar{\omega}\cos \theta +\eta \sin \theta \right)  
\nonumber \\
&&+\chi _{+}\left[ J_{x}^{2}\sin ^{2}\theta +J_{z}^{2}\cos ^{2}\theta +\sin
\theta \cos \theta \left( J_{x}J_{z}+J_{z}J_{x}\right) \right] .
\end{eqnarray}
Performing the contraction to the harmonic oscillator Hilbert space we find
the new Hamiltonian 
\begin{eqnarray}
F &=&\chi _{+}J\left( \frac{1}{2}\sin ^{2}\theta +J\cos ^{2}\theta \right)
-J\left( \eta \sin \theta +\Delta \bar{\omega}\cos \theta \right)   \nonumber
\\
&&+\left( a+a^{\dag }\right) \sqrt{\frac{J}{2}}\left[ \Delta \bar{\omega}%
\sin \theta -\eta \cos \theta -\chi _{+}\sin \theta \cos \theta \left(
2J-1\right) \right]   \nonumber \\
&&+\left( a^{2}+a^{\dag 2}\right) \chi _{+}\frac{J}{2}\sin ^{2}\theta
+a^{\dag }a\left\{ \eta \sin \theta +\Delta \bar{\omega}\cos \theta +\chi
_{+}\left[ J\sin ^{2}\theta -\left( 2J-1\right) \cos ^{2}\theta \right]
\right\}   \nonumber \\
&&+\left( a^{\dag }a^{2}+a^{\dag 2}a\right) \chi _{+}\sqrt{2J}\sin \theta
\cos \theta +a^{\dag 2}a^{2}\chi _{+}\cos ^{2}\theta .
\end{eqnarray}
Separating $F$ into Gaussian and non-Gaussian parts by imposing the
constraints in Eqs.~(\ref{eq:constraints}) gives

\begin{eqnarray}
F_{G} &=&\chi _{+}\left( \frac{J}{2}\sin ^{2}\theta +\cos ^{2}\theta \left(
J^{2}-\left\langle a^{2}\right\rangle ^{2}-2\left\langle a^{\dag
}a\right\rangle ^{2}\right) \right) -J\left( \eta \sin \theta +\Delta \bar{%
\omega}\cos \theta \right)  \nonumber \\
&&+\left( a+a^{\dag }\right) \sqrt{\frac{J}{2}}\left\{ \Delta \bar{\omega}%
\sin \theta -\eta \cos \theta +\chi _{+}\sin \theta \cos \theta \left[
-\left( 2J-1\right) +2\left\langle a^{2}\right\rangle +4\left\langle a^{\dag
}a\right\rangle \right] \right\}  \nonumber \\
&&+\left( a^{2}+a^{\dag 2}\right) \chi _{+}\left( \frac{J}{2}\sin ^{2}\theta
+\cos ^{2}\theta \left\langle a^{2}\right\rangle \right)  \nonumber \\
&&+a^{\dag }a\left( \eta \sin \theta +\Delta \bar{\omega}\cos \theta +\chi
_{+}\left\{ J\sin ^{2}\theta +\cos ^{2}\theta \left[ -\left( 2J-1\right)
+4\left\langle a^{\dag }a\right\rangle \right] \right\} \right) ,
\end{eqnarray}
and 
\begin{eqnarray}
F_{NG} &=&\left( a^{\dag }a^{2}+a^{\dag 2}a\right) \chi _{+}\sqrt{2J}\sin
\theta \cos \theta +a^{\dag 2}a^{2}\chi _{+}\cos ^{2}\theta  \nonumber \\
&&-\left( a+a^{\dag }\right) \chi _{+}\sqrt{2J}\sin \theta \cos \theta
\left( \left\langle a^{2}\right\rangle +2\left\langle a^{\dag
}a\right\rangle \right) -\left( a^{2}+a^{\dag 2}\right) \chi _{+}\cos
^{2}\theta \left\langle a^{2}\right\rangle  \nonumber \\
&&-a^{\dag }a4\chi _{+}\cos ^{2}\theta \left\langle a^{\dag }a\right\rangle
+\chi _{+}\cos ^{2}\theta \left( \left\langle a^{2}\right\rangle
^{2}+2\left\langle a^{\dag }a\right\rangle ^{2}\right) .
\end{eqnarray}
In these expressions, we have taken $\ \left\langle a^{2}\right\rangle
=\left\langle a^{\dag 2}\right\rangle $ which must be true by symmetry $%
\left( \left\langle \varphi \right\rangle =0\right) $.

\section{Effects of the non-Gaussian Hamiltonian}

\label{appen:wigner}

Here we show the details of the perturbation calculation to find the effects
of the non-Gaussian part of the Hamiltonian $F_{NG}.$ We show working only
for the first order correction. The second order calculation proceeds
identically but is much longer. We begin with the expression for the first
order perturbation to the Gaussian ground state 
\begin{equation}
\left| \Phi ^{\left( 1\right) }\right\rangle _{a}=\left| \tilde{0}%
\right\rangle _{a}+\sum_{k\ne 0}\frac{\mbox{\raisebox{-.7ex}{$_{a}$}}%
\!\left\langle \tilde{k}\right| F_{NG}\left| \tilde{0}\right\rangle _{a}}{%
E_{0}^{\left( 0\right) }-E_{k}^{\left( 0\right) }}\left| \tilde{k}%
\right\rangle _{a}.
\end{equation}
It is easier to work in the $b$ representation with the state 
\begin{equation}
\left| \Phi ^{\left( 1\right) }\right\rangle _{b}=\left| 0\right\rangle
+\sum_{k\ne 0}\frac{\left\langle k\left| F_{NG}\right| 0\right\rangle }{%
E_{0}^{\left( 0\right) }-E_{k}^{\left( 0\right) }}\left| 0\right\rangle ,
\label{eq:perturbed}
\end{equation}
where $F_{NG}$ must be expressed in the $b$ basis. Applying the Bogoliubov
transformation~(\ref{eq:bogul}) we obtain 
\begin{eqnarray}
F_{NG} &=&\chi _{+}\cos ^{2}\theta \left[ c^{2}s^{2}\left( b^{\dag
4}+b^{4}\right) -\left( c^{3}s+cs^{3}\right) \left( b^{\dag 3}b+b^{\dag
}b^{3}\right) +\left( c^{4}+s^{4}+4c^{2}s^{2}\right) b^{\dag 2}b^{2}\right] 
\nonumber \\
&&+\chi _{+}\sqrt{2J}\sin \theta \cos \theta \left[ \left(
cs^{2}-c^{2}s\right) \left( b^{\dag 3}+b^{3}\right) +\left(
c^{3}-s^{3}+2cs^{2}-2c^{2}s\right) \left( b^{\dag 2}b+b^{\dag }b^{2}\right)
\right]   \nonumber \\
&&+\Delta \left( b^{\dagger 2}+b^{2}\right) .  \label{eq:Fnonb}
\end{eqnarray}
where $c=\cosh r$ and $s=\sinh r,$ and $\Delta =T\left( c^{2}+s^{2}\right)
-csS$ accounts for any quadratic part left over from the self-consistent
approach. We have typically found this to be negligibly small. Substituting
Eq.~(\ref{eq:Fnonb}) in Eq.~(\ref{eq:perturbed}) we find the unnormalized
new state as 
\begin{equation}
\left| \Phi ^{\left( 1\right) }\right\rangle =k_{0}\left| 0\right\rangle
+k_{2}\left| 2\right\rangle +k_{3}\left| 3\right\rangle +k_{4}\left|
4\right\rangle ,
\end{equation}
with 
\begin{eqnarray}
k_{0} &=&1, \\
k_{2} &=&-\frac{\Delta }{\sqrt{2}\sqrt{S^{2}-4T^{2}}} \\
k_{3} &=&-\sqrt{\frac{2}{3}}\frac{\chi _{+}\sqrt{2J}\sin \theta \cos \theta
\left( cs^{2}-c^{2}s\right) }{\sqrt{S^{2}-4T^{2}}}, \\
k_{4} &=&-\sqrt{\frac{3}{2}}\frac{\chi _{+}\cos ^{2}\theta c^{2}s^{2}}{\sqrt{%
S^{2}-4T^{2}}}.
\end{eqnarray}
Setting the density matrix $\rho _{a}=\left| \Phi ^{\left( 1\right)
}\right\rangle _{aa}\left\langle \Phi ^{\left( 1\right) }\right| ,$ we
define the characteristic function 
\begin{eqnarray}
\chi \left( z\right)  &=&\text{Tr}\left\{ \rho _{a}\text{e}^{za^{\dag
}-z^{*}a}\right\}   \nonumber \\
&=&\text{Tr}\left\{ S^{\dag }\left( r\right) \rho _{a}S\left( r\right)
S^{\dag }\left( r\right) \text{e}^{za^{\dag }-z^{*}a}S\left( r\right)
\right\}   \nonumber \\
&=&\text{Tr}\left\{ \rho _{b}\text{e}^{b^{\dag }\left( zc+z^{*}s\right)
-b\left( zs+z^{*}c\right) }\right\}   \nonumber \\
&=&\sum_{\left\{ i,j\right\} \in \left\{ 0,2,3,4\right\}
}k_{i}k_{j}\left\langle i\left| \text{e}^{b^{\dag }\left( zc+z^{*}s\right) }%
\text{e}^{-b\left( zs+z^{*}c\right) }\right| j\right\rangle \text{e}^{-\frac{%
1}{2}\left( zc+z^{*}s\right) \left( zs+z^{*}c\right) },  \label{eq:charfun}
\end{eqnarray}
from which the Wigner function is found as 
\begin{equation}
W\left( \alpha \right) =\frac{1}{\pi ^{2}}\int \text{e}^{\alpha z^{*}-\alpha
^{*}z}\chi \left( z\right) \,\text{d}^{2}z.
\end{equation}
Expanding the exponential in the expectation value of Eq.~(\ref{eq:charfun})
and using the Rodrigues' formula for the Hermite polynomials $H_{n}\left(
x\right) =\left( -1\right) ^{n}\exp \left( x^{2}\right) d^{n}/dx^{n}\exp
\left( -x^{2}\right) $~\cite{abr72} one finds that the Wigner function has a
closed form expression as a sum of two-dimensional harmonic oscillator
functions. This makes for rapid numerical calculation, but the expression is
too lengthy to warrant inclusion.

\section{Inverse rotation of distributions on the sphere}

\label{appen:antirotate}

Suppose a contour ${\cal C}_{0}^{\text{p}}$ of the Wigner function in the
plane is parametrized as 
\begin{equation}
{\cal C}_{0}^{\text{p}}\left( t\right) =\left( x_{0}\left( t\right)
,y_{0}\left( t\right) \right) ,
\end{equation}
where $x=\alpha +\alpha ^{*},$ $y=-i\left( \alpha -\alpha ^{*}\right) $ are
quadrature variables. Projecting onto the sphere using the inverse of Eq.~(%
\ref{eq:mapping5}) we get 
\begin{eqnarray}
\theta _{0}\left( t\right) &=&c\sqrt{x_{0}^{2}\left( t\right)
+y_{0}^{2}\left( t\right) }, \\
\varphi _{0\,}\left( t\right) &=&\tan ^{-1}\left( \frac{y_{0}\left( t\right) 
}{x_{0}\left( t\right) }\right) ,
\end{eqnarray}
where care must be taken in determining the correct quadrant of $\varphi
_{0}.$ In Cartesian coordinates, the contour on the sphere is expressed 
\begin{equation}
{\cal C}_{0}^{\text{s}}=\left( j_{x}\left( t\right) ,j_{y}\left( t\right)
,j_{z}\left( t\right) \right) =J\left( \sin \theta _{0}\cos \varphi
_{0},\sin \theta _{0}\sin \varphi _{0},-\cos \theta _{0}\right) ,
\end{equation}
and is transformed by the rotation to 
\begin{equation}
{\cal C}_{1}^{\text{s}}=\left( j_{x}\cos \theta -j_{z}\sin \theta
,j_{y},j_{x}\sin \theta +j_{z}\cos \theta \right) ,
\end{equation}
which may then be reexpressed in terms of new spherical coordinates $\theta
_{1}$ and $\varphi _{1}$. Finally, if $\theta $ is small so that the number
of atoms in trap 2 greatly exceeds that in trap 1, we can obtain a Wigner
contour for the state of a ``single'' condensate by projecting the contour $%
{\cal C}_{1}^{\text{s}}\left( \theta _{1},\varphi _{1}\right) $ {\em directly%
} back to the plane using Eq.~(\ref{eq:mapping5}).

\begin{figure}[tbp]
\caption{Contours of the Wigner function on the Bloch sphere for exact
solutions with $N=100$ atoms and a) detuning $\Delta \bar{\omega}=0$,
nonlinearity $\chi_+=0.75$, b) $\Delta \bar{\omega}=30$, $\chi_+=0.75$, c) $%
\Delta \bar{\omega}=-0.44$, $\chi_+=0$. }
\label{fig:exactsphere}
\end{figure}

\begin{figure}[tbp]
\caption{Contours of the Wigner function projected into the plane for states
(a) and (b) in Fig.~1. The solid lines are the exact results. In (b), we
also show the prediction of the mean field approximation (dashed line), and
that using corrected versions of the quadratic moments (dot-dashed line). }
\label{fig:exactplane}
\end{figure}

\begin{figure}[tbp]
\caption{a)Mean angular position $\theta$ as a function of nonlinearity $%
\chi_+$ for $\Delta \bar{\omega} = 0,$ 25, 50, 75 and 100. b) Discrepancy in
the mean angle $\theta$ from the exact result as calculated by the mean
field picture (dotted) and corrected moments picture (solid). The curves are
labeled by the detuning $\Delta\bar{\omega}$. }
\label{fig:exacttheta}
\end{figure}

\begin{figure}[tbp]
\caption{Spread in the number difference $\delta n=(\mbox{Var}%
(n_1-n_2))^{1/2}$. Solid lines are exact results, dotted lines are
predictions of the corrected moments theory. The curves are labeled by the
detuning $\Delta\bar{\omega}$. }
\label{fig:numvar}
\end{figure}

\begin{figure}[tbp]
\caption{Exact states for negative nonlinearities. (a) Wigner function for $%
\chi_+=-0.01$, $\Delta \bar{\omega}=0$, (b) Wigner function for $%
\chi_+=-0.0115$, $\Delta \bar{\omega}=0$, (c) $\tilde{Q}$ function for $%
\chi_+=-0.012$, $\Delta \bar{\omega}=0$, (d) Wigner function for $%
\chi_+=-0.0115$, $\Delta \bar{\omega}=0.001$, }
\label{fig:negsqueeze}
\end{figure}

\begin{figure}[tbp]
\caption{Relative phase variance $\Delta \phi =\left( \left\langle
J_{y}^{2}\right\rangle -\left\langle J_{y}\right\rangle ^{2}\right) /(J/2)$
as a function of nonlinearity $\chi_+$. The legend indicates line types for
different detunings.}
\label{fig:phasesqueeze}
\end{figure}


\begin{references}
\bibitem{twotrapendnote1}  For example, the more complicated commutation
relations of angular momentum mean that the Bloch states are not the
eigenstates of a simple ``annihilation'' operator, and the $\tilde{Q}$
function defined in Eqs.~(19) and~(20) does not generate anti-normally
ordered moments of the operators $J_{-},J_{z}$ and $J_{+},$ as might be
hoped.
\end{references}
\end{document}